\documentclass[prd,showpacs]{revtex4}
\usepackage{caption}
\usepackage{amsmath}
\usepackage{graphicx}
\usepackage{epsfig}
\usepackage{dcolumn}
\usepackage{bm}
\usepackage{slashed}
\usepackage{amsfonts}
\usepackage{amssymb}
\usepackage[version=3]{mhchem}
\begin{document}
\title{\ Study of the magnetic anisotropy in a planar model of the La2CuO4 }
\author{Jos\'{e} C. Su\'{a}rez** and Alejandro Cabo* \textbf{,}}
\affiliation{$^{*}$\textit{Theoretical Physics Department, Instituto de Cibern\'{e}tica,
Matem\'{a}tica y F\'{\i}sica, Calle E, No. 309, Vedado, La Habana, Cuba }}
\affiliation{$^{**}$\textit{Facultad de F\'{\i}sica, Universidad de La Habana, Cuba.}}

\begin{abstract}
\noindent We study the anisotropy energy in a planar model for the $Cu-O$
layers of the \ce{La2CuO4} investigated in previous works. The Tight-Binding
(TB) potential of the model was extended out of the validity region with the
purpose of incorporate it in the potential. Next, the spin-orbit operator was
considered in the Hartree Fock solution of the full HF problem. It follows that in
spite of the fact that the extended potential possesses the square symmetry of
the crystalline structure, the anisotropy energy vanishes in its purely 2D
formulation. The result indicates that for the prediction by the model of the
observed non vanishing magnetic anisotropy at zero doping and temperature in
$La_{2}CuO_{4}$, its formulation requires of a 3D representation of its
Wannier orbitals and the use of $d$ like orbitals. The consideration of
more realistic $3D$ character of the model, by including multiple $CuO$
planes, also could be important due to the argued absence of long range order
in 2D.

 \bigskip
\end{abstract}
\maketitle

\section{Introduction}

The superconductivity is the intrinsic capacity of certain materials that allows to conduct electric currents without resistance and losses of energy in specific conditions. In 1911 it was discovery by Heike Kamerlingh Onnes\cite{onnes} when observed the electric resistance of the mercury disappear abrupt at 4.2K. In the next years several materials were discovered showing the occurrence of this phenomenon below certain critical temperature value $Tc$. A microscopic explanation was not arrived until 1960, when Bardeen, Schriefer and Cooper proposed a successful theory, today known as the BCS theory.

With the discovery of the $La_{2}CuO_{4}$ superconductor in 1986 at $Tc=30K$, was open a new stage , up to now devoted to the obtaining  and investigation of such kind of high $Tc$ superconductors. In this case, up to nowadays it had not been possible to develop a full explanation for high temperature superconductivity.

The essential characteristic of the first discovered HTc superconductor $La_{2}CuO_{4}$, and various others in its class, is that it has a crystalline structure containing CuO2 layers, inside a perovskite structure separated by block layers, who play a main role as charge reservoirs. Usually, the block layers are insulators and do not have any contribution to the low energy electronics states. On another hand in the \ce{CuO2} layers, the minimum energy of the electronics state are around the Fermi level.

At half filling (hole concentration p=0), the cuprates are antiferromagnetic insulators (AFI), with a Neel temperature near 300 $K$. When the number of holes grow, the antiferromagnetic phase is rapidly destroyed and the cuprates show a pseudogap phase. If you continue doping with hole, the metallic phase turns up.

In 1930 Bloch and Wilson \cite{ideta2} developed the band theory, that explain why several materials are metals and another are insulators. However the band theory failed in trying to explain the electronic structure of a large number of insulators. In this group were the \ce{NiO}, \ce{CoO} and \ce{La_{\mbox{\scriptsize{2-x}}}Sr_{\mbox{\scriptsize{x}}}CuO4}, which were
predicted as having metallic behavior. This result opened the doors for the profound study of the strongly correlated materials, which are characterized by showing strong correlation properties among the electrons. Two of the founding fathers of these studies Mott and Hubbard, attributed the insulator behavior to the electron-electron correlation.

Nowadays, a generalized criterion exists: For getting behaviors like the ones of insulators of Mott, it is necessary to take into account the short range correlations between electrons of opposite spin which are included in the phenomenological Hubbard models \cite{Ino2,Ino3}. It is also clear that the HF approximation, when is considered from first principles in a elemental electronic system, should not be able to predict the correlation between opposite spin electrons. In the recent works \cite{caboart,cabosimetria} by using a one band model, solved in the HF approximation, it was obtained a gap at $T=0$ for the \ce{La2CuO4}, then predicting its insulator behavior. This reflect that some of the important properties of these materials, that are products of the strong electronic correlations, can be predicted by the model in the HF approximation. This outcome does not result strange, after taking in mind that, by example the antiferromagnetism, that is a strong correlation property, can be derived form a HF solution of the Hubbard model. The main point to note in this, is that the considered  model is not the original first principle electronic system. In fact the model only differs from a Hubbard one in that the near neighbor approximation was not taken.

Since the Hamiltonian adopted in the model does not includes the spin, the HF solution do not determine a direction for the antiferromagnetism. This fact define the basic motivation of this work. It consists in including the spin-orbit interaction in the starting Hamiltonian and then study the effect of the spin-orbit interaction in the solutions. The expectation was to study the magnetic anisotropy of the model. Therefore, he specific objectives of our work are the following ones:

\begin{enumerate}
  \item Obtain the form of the spin-orbit operator in the frame of the mentioned model for the superconductor material \cite{caboart,cabosimetria}.
  \item Afterwards, to evaluate the anisotropy energy of the antiferromagnetic HF states of the model, in the no doping limit of the \ce{La2CuO4}. For this purpose the mean value of the spin-orbit operator of the electrons with the crystalline field \cite{caboart,cabosimetria} is evaluated in the HF states. Since the HF solution is degenerated with respect to spin rotations, to evaluate the mean value, the HF states were arbitrarily rotated in the spin structure of their many particles \cite{caboart,cabosimetria}, by unitary transformations. These spin rotations are the ones representing corresponding space rotations around an arbitrary spatial axis.

\end{enumerate}

\bigskip

\section{\ce{La2CuO4} System}

\begin{figure}[ht]
\begin{center}
  \includegraphics[scale=.13]{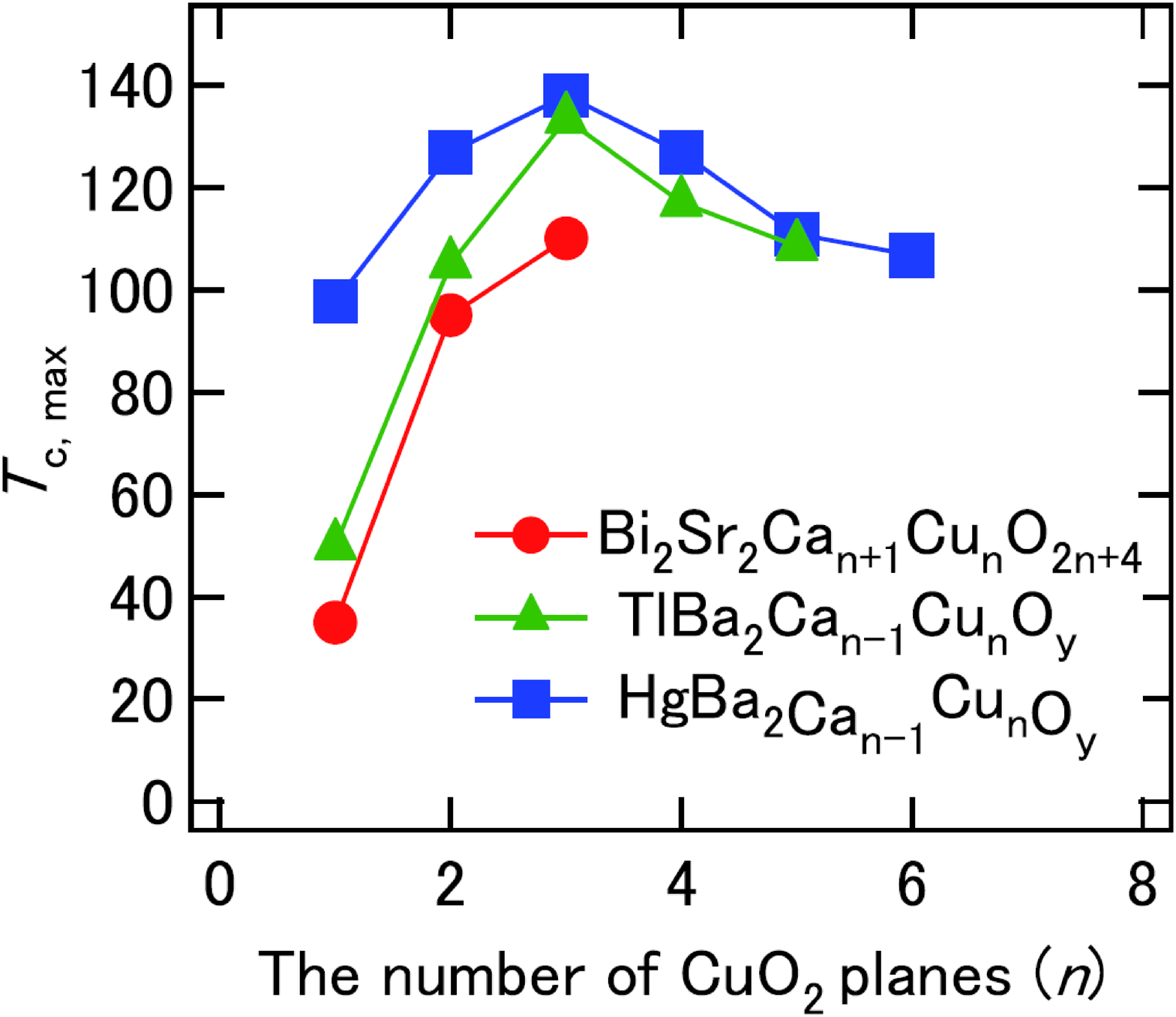}\\
\end{center}
  \caption{Dependence of Tc of the HTSC with the number of serial planes \ce{CuO2}.}\label{fig_Noplanos_sc}
\end{figure}

In general, the HTSCs presents a tetragonal structure, and have one or more planes \ce{CuO2} in its structure separated by layers of another atoms (Ba, O, La,...). In the \ce{CuO2} planes, each cooper ion is coupled to four oxygen ions separated an approximate distance of 1.9 {\AA }. The critical temperature depended of the maximum number of adjoint layers \ce{CuO2} (figure
\ref{fig_Noplanos_sc}). The electronic behavior of the cuprates is very strongly determined by the electronic density in the two dimensional layers \ce{CuO2}. The basic behavior of the \ce{CuO2} layers, is common for all the cuprate superconductors.

\subsection{Crystalline structure and phases diagrams}

\begin{figure}[ht]
\begin{center}
  \includegraphics[scale=.20]{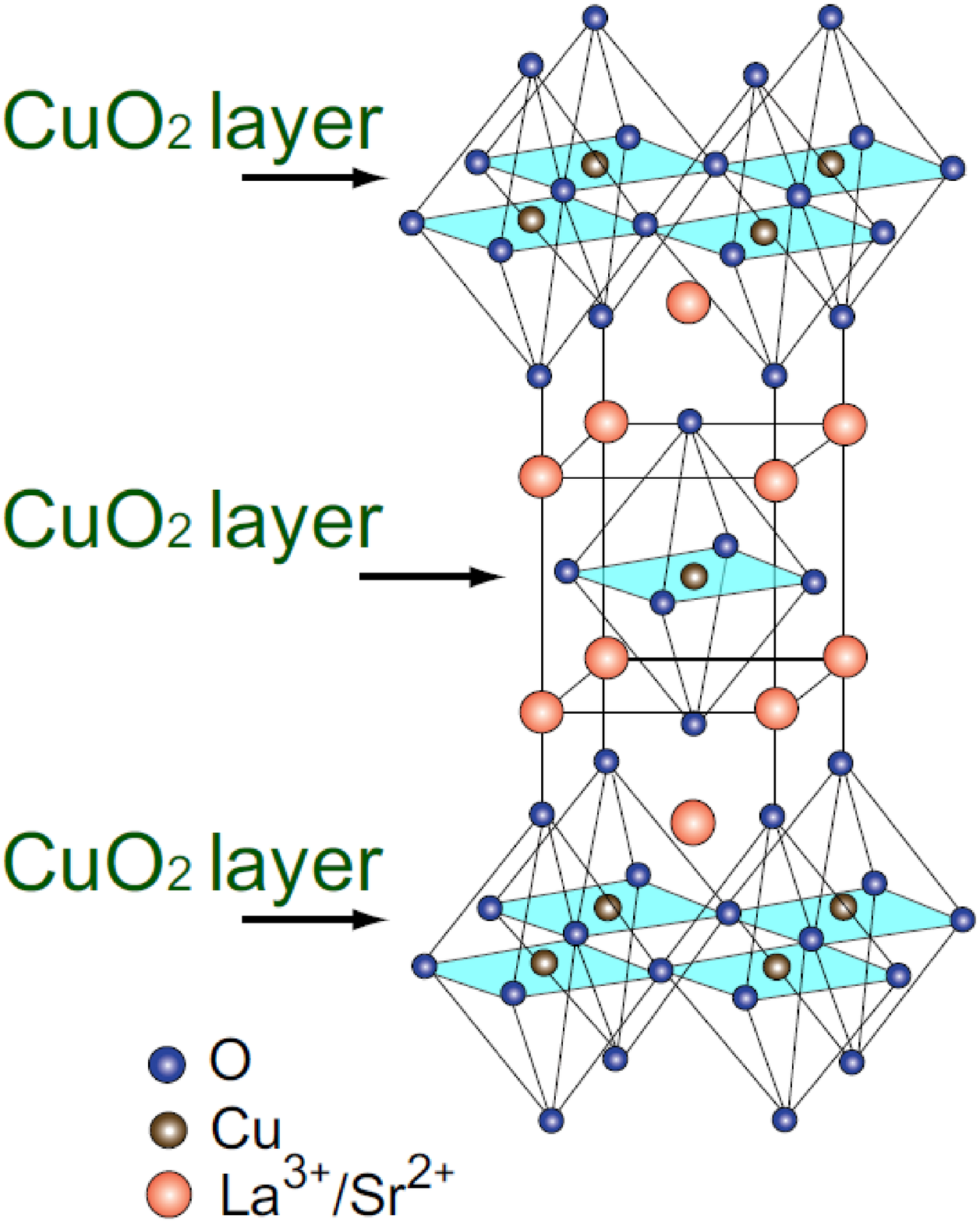}\\
\end{center}
  \caption{Crystalline structure of the \ce{La2CuO4}}\label{fig_htc_ec}
\end{figure}

The family of monolayer compounds, \ce{La_{\mbox{\scriptsize{2-x}}}Sr_{\mbox{\scriptsize{x}}}CuO4}, where x indicates the magnitude of strontium (\ce{Sr}) doping, crystallize to the
tetragonal structure centered in body (bct), usually named T structure, showed in the figure \ref{fig_htc_ec}. In the \ce{La_{\mbox{\scriptsize{2-x}}}Sr_{\mbox{\scriptsize{x}}}CuO4} the planes \ce{CuO2} are separated by $\thicksim$ 6.6 {\AA }, separated by 2 planes of \ce{LaO}, that act like charge reservoirs, absorbing electrons of the conductor planes under the doping of the material with holes.

Exist consensus in consider the electronic structure of the \ce{La_{\mbox{\scriptsize{2-x}}}Sr_{\mbox{\scriptsize{x}}}CuO4}, as quasi two dimensional. This low dimensionality of this electronic system allows to be treated from another viewpoint. For example collective phenomena, such as fluctuations in spin density waves (S.D.W) and charge density waves (C.D.W), could be the origin of the superconducting mechanism \cite{cabo4,cabo5}.

\begin{figure}[ht]
  \begin{center}
  \includegraphics[scale=.23]{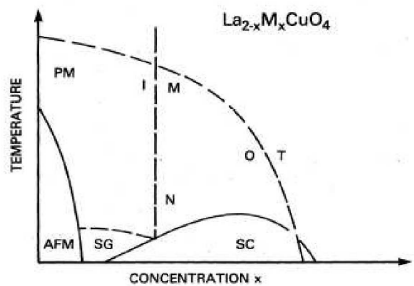}\\
\end{center}
   \caption{Phases diagram of the \ce{La_{\mbox{\scriptsize{2-x}}}M_{\mbox{\scriptsize{x}}}CuO4}. AF=Antiferromagnetic, PM=Paramagnetic, I=Insulator, M=Metal, SG=SpinGlass, N=Normal, SC=Superconductor, T=thetragonal structure, O=Orthorhombic structure.}\label{fig_tran_fase1}
\end{figure}

In the figure \ref{fig_tran_fase1}, it is illustrated the phase diagrams in the (x,T) plane for the compound \ce{La_{\mbox{\scriptsize{2-x}}}Sr_{\mbox{\scriptsize{x}}}CuO4}, where \ce{T}
is the temperature. Even when the absolute antiferromagnetic order and the insulator Mott state persist for a wide range of temperatures, it is very sensible at doping. For example, in the $T=0$ limit, it is lost at $x=0.04$ doping, but in the pseudogap (SG) phases and superconductor (SC) ones, exist experimental evidence of a less extended AFM order \cite{cabo2,cabo6}. All the ideas mentioned before support the possibility that the AFI state (x=0) play a crucial role in guaranteeing first: a correlation between pairs of holes (SG), second: a condensation of these holes toward the SC state \cite{cabo7}.

In this compound a structural phase transition occurs. At high temperatures the structure is tetragonal, but for low temperature the Cu atoms and the six oxygens around them, stray slowly of their positions, forming an orthorhombic structure. In the majority of the theoretical studies about this compound, this little distortion is usually ignored.

The crystalline structure of the LSCO have simple \ce{CuO2} planes, while the systems \ce{Bi2Sr2CaCu2O_{8+y}} and \ce{YBa2Cu3O7} (YBCO) has double \ce{CuO2} planes. The intermediate layers of LSCO do not have the \ce{Cu-O} chains that affects considerably the \ce{CuO2} planes, in contrast with YBCO, and the complicate structure that present the intermediate layers of the \ce{Bi2212}.

\section{The starting theoretical model}

\begin{figure}[ht]
  \begin{center}
  \includegraphics[scale=.40]{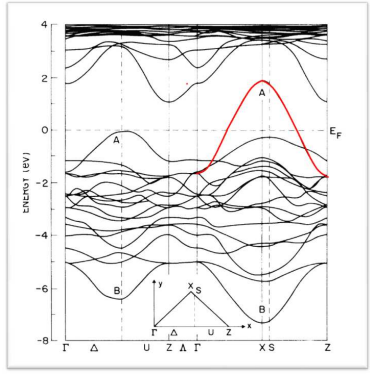}\\
\end{center}
  \caption{Bands structure for the \ce{La2CuO4} calculated by Horsch y Stephan et.al. 1993. The semifilled band $\Gamma-X$ predict a tight-binding behavior in the reciprocal plane}\label{fig_Banda_matheiss}
\end{figure}

The band diagram shown in the figure \ref{fig_Banda_matheiss} associated to
the \ce{La2CuO4} was obtained by means of techniques of LAPW\cite{cabo32}. The
last occupied band is half filled, predicting a metallic behavior, and its
form suggests a Tight-Binding nature for the electron gas that occupied it
within the effective environment of the rest of the bands. The less bound
electrons to the compound \ce{La2CuO4}, are the ones of the \ce{Cu2+} atoms,
that do not have closed their last shells (3d) , at difference with the
situation with the \ce{O2-} ions. These electrons are the ones that form the
last band in the corresponding solid.

The model proposed in the references \cite{caboart,cabosimetria} is described
in this section. One of the most used pictures to describe the movement of the
valence electrons in a solid with a periodic potential is the tight-binding
model. This model is applicable when the overlapping of the atomic orbitals
of the neighbors can be considered small and it can be supposed that is not
necessary to carry out significant corrections to the atomic wave functions,
created by the interaction with the neighboring orbitals. The tight-binding
approximation provides a reconciliation among the seemingly contradictory
representations of the very localized atomic levels and on another hand the
free electron model \cite{ponce9}, in which the wave functions are lineal
combination of planes waves.

\begin{figure}[ht]
\begin{center}
  \includegraphics[scale=.25]{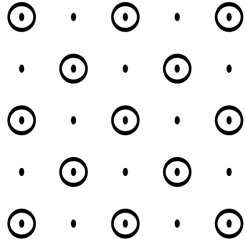}\\
\end{center}
  \caption{Punctual lattice associated to the \ce{CuO} planes. To obtain the AFM properties of the electronic gas, and more general, to release the symmetries restriction, it is separated the lattice in the 2 represented red} \label{fig_Sub_redes}
  \end{figure}

From the previous comments we can suppose that these electrons are strongly
correlated to the base cells \ce{CuO2}, preferably toward the corresponding
centers of \ce{Cu} \cite{cabo18}. Then, the lattice will be supposed as the
square lattice of $Cu$ atoms located at the $CuO$ planes (see figure
\ref{fig_Sub_redes}).

The electrons not belonging to the last band and the nuclear charge that
neutralizes them, play a double role. They act in one hand, as an effective
polarizable medium that screens the field produced by a point charge external
to them, and in second place, due to their spatial distribution and magnitude,
they guarantee with their action the periodic order of the solid, that is
modeled with a periodic potential in the lattice of points.

The Hamiltonian of the model introduced in the references take the form:

\begin{equation}\label{eqn_hamTotal}
  \hat{H}= \sum^{N}_{i=1} \frac{p_{i}^{2}}{2m} + \sum^{N}_{i=1} W_{\gamma}(\vec{x}_{i}) + \sum^{N}_{i=1}F_{b}(\vec{x}_{i}) + \frac{1}{2}\sum^{N}_{i=1}\sum^{N}_{j\neq i}V(\vec{x}_{i},\vec{x}_{j}) + \sum^{N}_{i=1}\hat{U}_{SO}
\end{equation}

where  $\frac{\hat{p}_{i}^{2}}{2m}$ is the kinetic energy term,
$W_{\gamma}(\vec{x}_{i})=W_{\gamma}(\vec{x}_{i}+\vec{R})$ is the periodic potential that exercise the electrons in the partially filled single band, and $\vec{R}$ runs over the points of the lattice of \ce{Cu} (see figure \ref{fig_Sub_redes}),

$F_{b}(\vec{x_{i}})=\sum_{\vec{R}}\int d^2y\frac{e^2}{4\pi\epsilon\epsilon_{0}}\frac{\exp(-\frac{\vec{y}-\vec{R})^{2}}{b^{2}})/\pi b^2}{|\vec{x_{i}}-\vec{y}|},\ b\ll p.$ is the potential of the jellium background correspondingly only to the surplus of charges which neutralize the electrons that partially filled the band,

$V(\vec{x_{i}},\vec{x_{j}})=\frac{e^2}{4\pi\epsilon\epsilon_{0}}\frac{1}{|\vec{x_{i}}-\vec{x_{j}}|}$ screened Coulomb potential,

$V(\vec{x_{i}})=W_{\gamma}(\vec{x}_{i})+F_{b}(\vec{x_{i}})$ is the effective potential that feel the electrons that partially filling the last band,

$\hat{U}_{SO}=\frac{\hbar}{2m^{2}c^2} \nabla V(\vec{x_{i}})\cdot\vec{\sigma}\times\hat{p}$ is the spin-orbit interaction term. \\

\subsection{The Hartree-Fock method} 

One of the most useful techniques to find approximate solutions to the
Schrodinger%
\'{}%
s equation is the molecular orbitals approximation. In this approach it is
considered that the electrons occupy spin-orbital functions of a given energy.
This technique is based in the \emph{variational theorem}, which state that the
expected value of $\hat{H}$ for an arbitrary wave function is not smaller that
the lowest eigenvalue $E_{0}$ corresponding to the exact solution of the
equation\cite{jensen}. Then, the best wave function is obtained when the
energy calculated by the equation (\ref{variacHF}) has a minimum.%

\begin{equation}\label{variacHF}
    E=\frac{\langle\Psi|\hat{H}|\Psi\rangle}{\langle\Psi|\Psi\rangle}.
\end{equation}

Then, the problem is reduced to find the solution by finding it form in all the
allowed many electronic states, under the condition of making minimal the energy of the system. In the HF formalism one keeps each electron in a spin-orbital, in the
field of the nucleus and the other electrons that are in their respective  spin-orbitals
\ (fixed), subject to the constraint that the spin-orbitals are orthonormal.
After this minimization process, the HF equations for individual
spin-orbitals result:%

\begin{equation}\label{fock}
    \hat{f}_{i}\phi_{i}=\varepsilon_{i}\phi_{i},
\end{equation}

\noindent where $\hat{f}_{i}$ is the Fock operator and $\varepsilon_{i}$ are the
eigenvalues that represent the energies of the spin-orbitals ($\phi_{i}$).

The Fock operator for an arbitrary electron $^{\prime}1^{\prime}$ (equation
\ref{operFock}) can be separated in three terms like it is shown next:%

\begin{equation}\label{operFock}
    \hat{f}_{i}(1)=\hat{H}^{0}(1)+\sum_{i=1}^{N}[\hat{J}_{i}(1)-\hat{K}_{i}(1)],
\end{equation}

where the Hamiltonian operator $\hat{H}^{0}(1)$ is defined by:

\begin{equation}\label{coreFock}
    \hat{H}^{0}(1)= \frac{p_{1}^{2}}{2m} + W_{\gamma}(\vec{x_{1}}) + F_{b}(\vec{x_{1}}).
\end{equation}

Here and in what follows, the arguments of the wave functions $^{\prime}1^{\prime}$ and
$^{\prime}2^{\prime}$, indicate the collection formed by the spacial
coordinates and the spin. In absence of interelectronic interactions
(\ref{coreFock}) will be the only operator corresponding to the movement of a
single electron in the field of the nucleus and the other electrons, \ that
\ do not pertain to the partially filled band crossing the Fermi level.

The direct interaction operator is

\begin{equation}\label{coulFock}
    \hat{J}_{j}(1)\phi_{a}(1)=\left\{\int d\tau_{2}\phi_{j}(2)\frac{e^{2}}{4\pi\epsilon\epsilon_{0}r_{12}}\phi_{j}(2)\right\}\phi_{a}(1),
\end{equation}

\noindent where $\int d\tau_{2}$ represents the integral for the space coordinates and
the sum for the spin coordinates. This operator correspond to the average
potential of an electron in the state $\phi_{j}$.

On the other hand the exchange operator is expressed as:

\begin{equation}\label{excFock}
    \hat{K}_{j}(1)\phi_{a}(1)=\left\{\int d\tau_{2}\phi_{j}(2)\frac{e^{2}}{4\pi\epsilon\epsilon_{0}r_{12}}\phi_{a}(2)\right\}\phi_{j}(1).
\end{equation}

The equation (\ref{fock}) is known as the canonical HF equation and the fact
that it corresponds to a non linear eigenvalue and eigenfunctions \ problem,
implies that to find the solutions becomes necessary an iterative procedures.
This fact leads to propose a group of test spin-orbitals that allow to obtain
a new group of such orbitals, from which a new Fock operator is generated. The
process is repeated until the energy and the functions converges, such a
procedure is named the self-consistent field process.

\subsection{ The approximate basis functions }

In a general way the spin-orbitals ($\phi_{i}$) are expanded by means of a
well-known group of basis functions associated with the geometric positions of
the nuclei. This basis should satisfy two practical conditions

\begin{enumerate}
  \item Should reasonably describe the physics of the problem.
  \item The integrals that appears in the mathematical treatment should be
evaluates with relative easiness.
\end{enumerate}

The spatial components of the spin-orbitals is expanded in terms of the basis
functions generally centered in atoms, according to (\ref{baseDes}). The
expansion which must be infinite in principle, is truncated using a number of
functions \textbf{M}, being $c_{\nu i}$ the coefficients determining each
atomic orbital contribution $\chi_{\nu}$ in the linear expansion
\cite{Roothaan}

\begin{equation}\label{baseDes}
    \phi_{i}=\sum_{\nu=1}^{M} c_{\nu i}\chi_{\nu}.
\end{equation}

Substituting the orbitals expanded according to (\ref{baseDes}) in the HF
equation \cite{Pimentel}, a set of matrix equations is obtained which can be:

\begin{equation}\label{scf}
    \mathbf{FC}=\mathbf{SC\varepsilon},
\end{equation}

\noindent where \textbf{F} is the Fock matrix for the $F_{\mu\nu}=\langle\chi_{\mu
}|f|\chi_{\nu}\rangle$; \textbf{S} is the superposition matrix among the
elements of the base $S_{\mu\nu}=\langle\chi_{\mu}|\chi_{\nu}\rangle$ and
\textbf{C} is the coefficients matrix $c_{\nu i}$.

In the beginnings, the developing of HF techniques for systems of many
particles, like bands and polyatomic complex calculations, was limited by the
memory and speed of the machine calculators of that times. It was necessary to
devise, on the basis of arguments and credible approaches, for example like
the hole%
\'{}%
s Fermi model \cite{slater}, mean field approximation, etc, a reduction of
the functional space for searching and of the complications of the
calculations. An useful assumption is to consider that the solution states of
the system of equations(\ref{scf}) have quantized spin in the same direction
in all the space, that is to say:

\begin{eqnarray} \label{e_alphabeta}
\phi_{\textbf{k}}(x,s)=\begin{cases}\phi^{\alpha}_{\textbf{k}}(x)\
u_{\uparrow}(s) & \text{ type $\alpha$ state},\\
\phi^{\beta}_{\textbf{k}}(x)\ u_{\downarrow}(s) & \text{
type $\beta$  state}.
\end{cases}
\end{eqnarray}

If the spatial functions $\phi_{\mathbf{k}}^{\alpha}$ and $\phi_{\mathbf{k}%
}^{\beta}$ are identical, the HF calculations is named as restrictive,
otherwise unrestricted \cite{szabo}. However, due to their definitions, both
cases are restrictive.

It can be said that the HF\ system of equations (\ref{scf}) is written in a
form being invariant under rotation, because it is not considered the spin
quantization in one absolute direction. That is, the states to be found are
not $\alpha$ type neither $\beta$ one. The first in derive the HF equations in
this fully unrestricted form was P. A. M. Dirac in the 1930 \cite{dirac}. Work in this rotation invariant way wide the functional space being investigated. In consequence it allows to
extend the HF scheme towards the search of \ ground states showing magnetic
character. This possibility is suggested in references
\cite{caboart,cabosimetria}. These works considered a base that allowed to
find HF solutions for the spin-orbitals showing spin quantization depending of
the spatial position.

The Tight-Binding basis in the one band approximation, that was used in these
works consists of 4 sets of basis functions:

\begin{eqnarray}\label{base de Bloch}
\varphi^{(r,\sigma_z)}_{\textbf{k}}(\textbf{x},s) &=& \sqrt{\frac{2}{N}}u^{\sigma_z}(s)\sum_{\textbf{R}^{(r)}}\exp(i\textbf{k}\cdot\textbf{R}^{(r)})\varphi_0(\textbf{x}-\textbf{R}^{(r)}),\\
\hat{\sigma}_zu^{\sigma_z} &=& \sigma_zu^{\sigma_z},\\
\varphi_0(\textbf{x}) &=& \frac{1}{\sqrt{\pi a^2}}\exp(-\frac{\textbf{x}^2}{2\ a^2}),
\end{eqnarray}

where $a\ll p$, $r=1,2$ and $\sigma_z=\pm\frac{1}{2}$.

The equation (\ref{scf}) projected in the tight binding basis
(\ref{base de Bloch}) was written in the form:

\begin{eqnarray}\label{EcuMatricial}
[\textrm{H}_{\textbf{k}}+ \widetilde{\chi} \ (
\textrm{J}_{\textbf{k}}-\textrm{K}_{\textbf{k}}-\textrm{F}_{\textbf{k}})]\textbf{.}\textrm{B}^{\textbf{k},l}=
\widetilde{\varepsilon}_{l}(\textbf{k}) \
\textrm{S}_{\textbf{k}}\textbf{.}\textrm{B}^{\textbf{k},l},
\end{eqnarray}

where ${H}_{\mathbf{k}}$ is Hamiltonian part that corresponds to the kinetic
energy and the total effective potential,

$\textrm{J}_{\textbf{k}}$ is the direct interaction term,\\

${K}_{\textbf{k}}$ is the exchange term,\\

${F}_{\textbf{k}}$ it is the neutralizing bottom potential,\\

${S}_{\textbf{k}}$ is the overlapping matrix.\\

Also $\widetilde{\chi}\equiv\frac{me^{2}a}{4\pi\hbar^{2}\epsilon\epsilon_{0}%
}\frac{a}{p}$ and $\widetilde{\varepsilon}_{l}(\mathbf{k})\equiv\frac{ma^{2}%
}{\hbar^{2}}\varepsilon_{l}(\mathbf{k})$ are the dimensionless parameters in
those that $a$ is the characteristic radius of the functions $\varphi_{0}$,
$p=3.8{\mathring{A}}$ is the separation between two $Cu$ near neighbors in the
absolute red, $\hbar$ is Planck constant, $m$ and $e$ are the mass and the
charge of the electron respectively.

In the low overlapping approximation among nearest neighbors (that is to say
among different sublattices) the functions of the base that correspond to
different sublattices with the same spin quantization are the only one, not
being orthogonal. However, the orthogonalization of different elements of the
same sublattice, as well as the unit norm of all elements, are valid. The
$\varphi_{0}(\mathbf{x}-\mathbf{R}^{(r)})$ orbitals will be called as Wannier
orbitals and represent the probability amplitude to find an electron in the
surroundings of a point of the lattice.

\section{Magnetics properties} 

Magnetism arises on the sub-atomic level from localized polarization of the
electron clouds of certain atoms arising from unpaired electrons. This causes
the charge on the atom to have a lattice angular momentum. Any flow of charge
causes additional physical effects on the surroundings, usually referred to as
a $magnetic$ effect. In the case of atoms the lattice angular momentum of the
charge cloud causes a magnetic field perpendicular to the rotation plane of
the charge. The magnitude of this magnetic or spin moment is dependent on the
species of atoms \cite{york3}. How these spin moments interact with each other
is critical to how different materials are characterized magnetically. When
atoms are brought in proximity to each other there is a probability of an
electron jumping from one atom to another, known as the Heisenberg exchange
[4]. This interaction probability can indirectly couple the spin moments of
the atoms, causing the spin moments to align parallel or anti-parallel. In
most materials the spin moments are small and aligned randomly, leading to
paramagnetism as shown in Figure. In some materials however, specifically
transition metals such as nickel, cobalt, and iron, the spin moments are
large, and align in parallel. This causes a lattice spontaneous magnetic
moment in the material.

\begin{figure}[ht]
\par\begin{centering}
{\scriptsize
(a)}\includegraphics[scale=0.25]{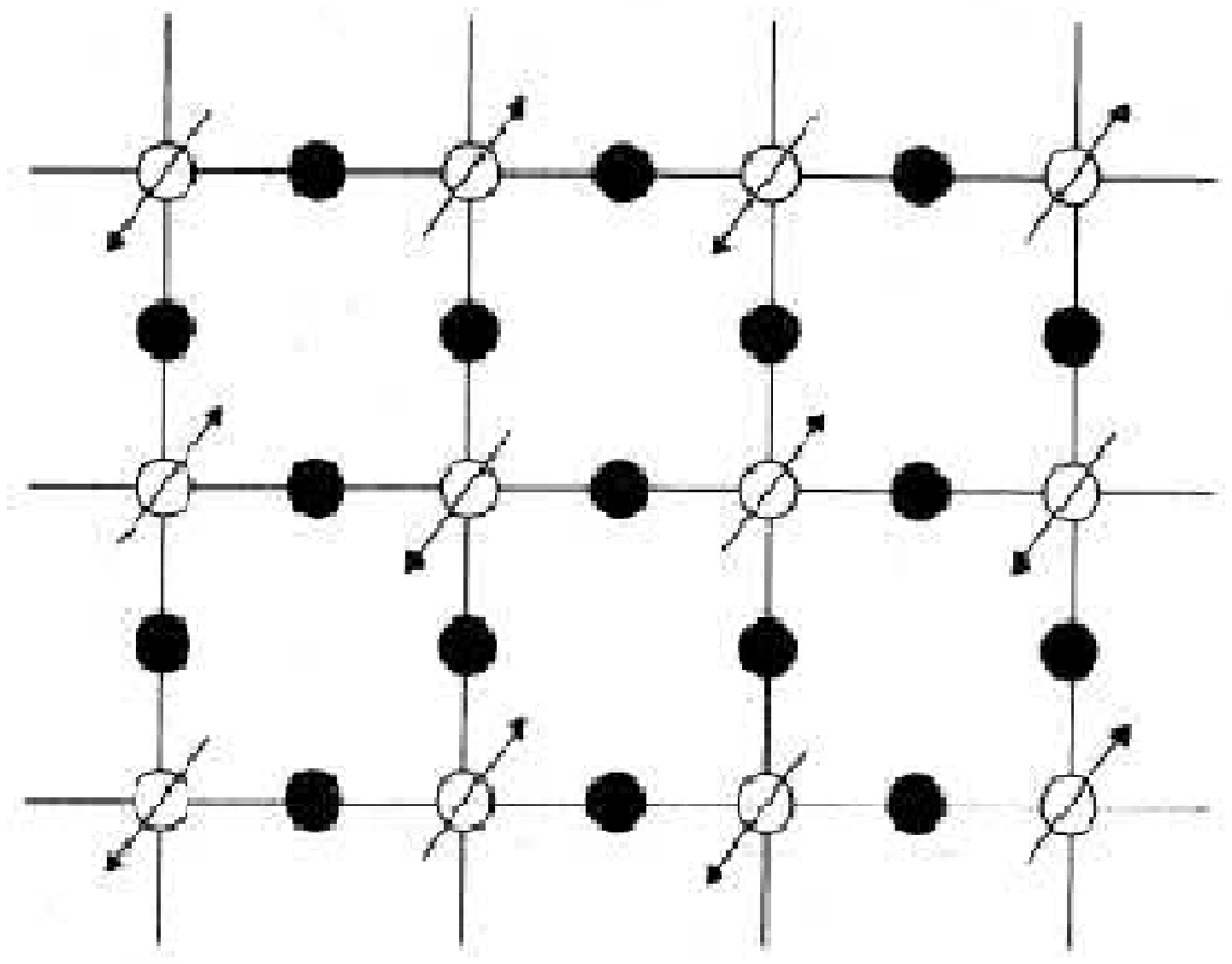}\hspace{8mm}{\scriptsize
(b)}~~\includegraphics[scale=0.25]{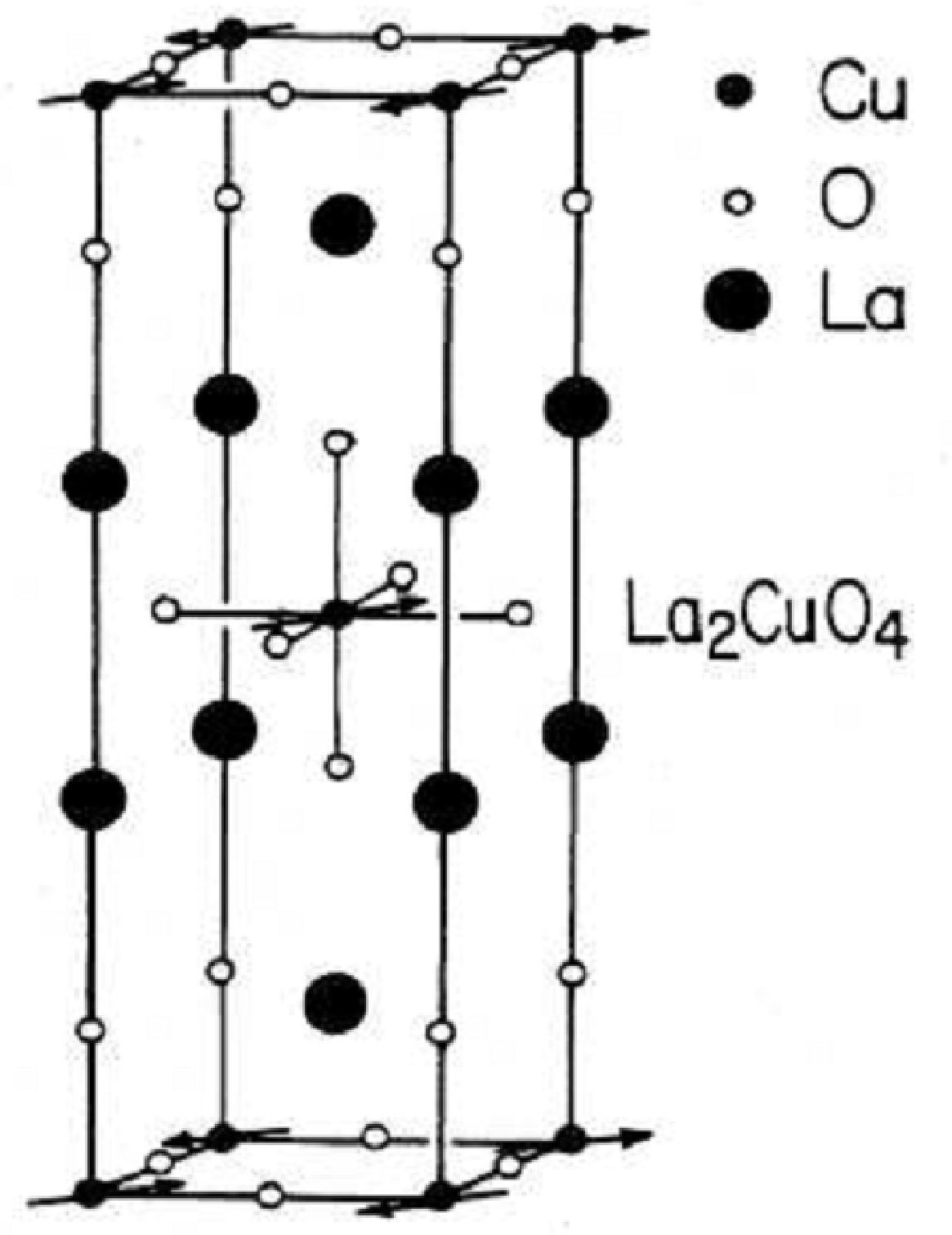}
\par\end{centering}
\caption{(a) Magnetic structure of the \ce{La_{\mbox{\scriptsize{2-x}}}Sr_{\mbox{\scriptsize{x}}}CuO4}. Schematic representation of the \ce{CuO_{2}} planes, the cooper and the oxygen are represented by means of  hollow and padded circles, respectively (b) Magnetic structure of the \ce{La_{\mbox{\scriptsize{2-x}}}Sr_{\mbox{\scriptsize{x}}}CuO4} in its planes \ce{CuO2}}%
 \label{fig_spin_plan3D}
\end{figure}

In \ce{La_{\mbox{\scriptsize{2-x}}}Sr_{\mbox{\scriptsize{x}}}CuO4} case, the
use of scattering of neutrons techniques \cite{cabo8}, has contributed
important information about the magnetic structure of these materials. In the figure \ref{fig_spin_plan3D} the AFM profile of the
\ce{La2CuO4} in the copper-oxygen is shown. The places \ce{Cu} contains the
magnetic moment, whose address rests approximately to 45 grades on the link
\ce{CuO}. It happens that the ions \ce{O2-} and \ce{La3+} complete their more
external layers, while the \ce{Cu2+} lacks an electron for it. Then the ion
\ce{Cu} presents an electron without matching up. It is also the less bound
one to the arrangement.

\section{\bigskip}

\section{Anisotropy energy and magnetocrystalline anisotropy}

One of the most basic parameters in a magnetic system is the
magneto-crystalline anisotropy, that is, the preference of the spin moments to
align with some fixed crystallographic axes. This is a consequence of the
interaction of the local environment though its spin-orbit interaction with
the electrons. The simplest form of anisotropy is the uniaxial one that
consists in that the magnetic moment prefers to align throughout a simple axis
$\vec{e}$, usually called the easy axis.

Considering the operator $\sum_{i=1}^{N}\hat{U}_{SO}$ as an perturbation of
the initial Hamiltonian \ref{eqn_hamTotal}, the anisotropy energy will be
evaluated in this subsection, in the first correction to the HF energy of the
system, according to the corresponding formula of perturbation theory.

\begin{equation}\label{eqn_EnergAnis}
    E_{\text{anisotropy}}(\vec{n},\theta)=\sum_{k,l}<\phi^{R}_{k,l}|\hat{U}_{SO}|\phi^{R}_{k,l}>,
\end{equation}

\noindent where the $|\phi_{k,l}^{R}>$ orbitals are obtained of rotating the orbital
solution $|\phi^{k,l}>$ around the axis defined by the unitary vector $\vec
{n}$ in an angle $\theta$. To rotate the orbital one should interpret the
following procedure: Let us suppose that we have an orbital that is
eigenfunctions of the operator $\hat{\sigma_{z}}$ and we make a rotation
around the axis defined by the unitary vector $\vec{n}$ in an angle $\theta$.
Then, the $z$ axis will become an $z%
\acute{}%
$ axis, and the orbital will be transformed in an orbital that is
eigenfunction of $\hat{\sigma_{z^{\prime}}}$ according to (\ref{eqn_orbrot}).

\begin{align}\label{eqn_orbrot}
|\phi_{k,l}>=\sum^{r,\sigma_z}B_{r,\sigma_z}^{k,l}|\varphi^{(r,\sigma_z)}_{\textbf{k}}>,\\
|\phi^{R}_{k,l}>=U(a,b)|\phi_{k,l}>, \\
U(a,b)= (\begin{array}{cc}
a & b\\
-b^{*} & a^{*}\end{array}),\\
a=cos(\frac{\phi}{2})-i n_{z}\sin(\frac{\phi}{2}),\\
b=-n_{y}\sin(\frac{\phi}{2})-i n_{x}\sin(\frac{\phi}{2}).
\end{align}

A rotation of the orbital leads to a rotation of the antiferromagnetic
structure of the material at $T=0$. This can be expressed in the following way
\cite{sakurai}:

\begin{eqnarray}
  \vec{M}_{R}(\vec{x}) &=& \sum_{k,l}<\phi^{R}_{k,l}(\vec{x})|\vec{\sigma}|\phi^{R}_{k,l}(\vec{x})>, \\
  \vec{M}_{R}(\vec{x}) &=& \vec{\vec{R}}\sum_{k,l}<\phi_{k,l}(\vec{x})|\vec{\sigma}|\phi_{k,l}(\vec{x})>, \\
  \vec{M}_{R}(\vec{x}) &=& \vec{\vec{R}}\vec{M}(\vec{x}).
\end{eqnarray}

\noindent where $\vec{M}_{R}(\vec{x})$ is the magnetization vector rotated around an
$\vec{n}$ axis, in an angle $\theta$ starting from the initial direction of
the magnetization vector $\vec{M}(\vec{x}),$ and $\vec{\vec{R}}$ is the tensor
that rotates the magnetization vector in the three-dimensional space.

\begin{figure}[ht]
\begin{center}
  \includegraphics[scale=.25]{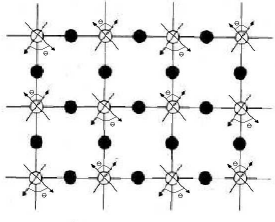}\\
\end{center}
  \caption{Antiferromagnetic structure in the \ce{CuO} planes rotated around the z axis in an angle $\theta$.}\label{fig_anti_rotada}
\end{figure}

The quantity $E_{\text{anisotrop-a}}(\vec{n},\theta)$ can be interpreted in
the following way: Let us suppose that the system has its antiferromagnetic
structure in certain direction and that this structure is rotated around the
axis $\vec{n}$ in an angle $\theta$. Then $E_{\text{anisotrop-a}}(\vec
{n},\theta)$ is the correction to the HF energy of the system produced by the
spin-orbit interaction, when the magnetic structure has that direction (see
figure \ref{fig_anti_rotada}).

\section{Anisotropy energy with Gaussian Wannier orbitals}

The anisotropy energy was studied in the initial model. That is, considering
as the basis functions, the Bloch functions constructed in the sublattices by
means of defined by Wannier Gaussian orbitals. It was already seen that in
this two-dimensional model the spin-orbit operator reduces to the following form:

\begin{equation}\label{eqn_SO}
  \hat{U}_{SO}= -\frac{\hbar}{2m^{2}c^2}\hat{\sigma_{z}}.(E_{x}\hat{p}_{y}-E_{y}\hat{p}_{x}).
\end{equation}

In the following developments, it will be considered that $g=-\frac{\hbar
}{2mc^{2}}$, where $m$ is the electron mass.

The expression for the energy to evaluate is given for:%

\begin{equation}\label{eqn_EnergAnis4}
    E(\vec{n},\theta)=\sum_{k,l}<\phi^{R}_{k,l}|\hat{U}_{SO}|\phi^{R}_{k,l}>,
\end{equation}

\noindent where $B_{r,\sigma_{z}}^{k,l}$ are the constants for the expansion of the HF
solutions of the original problem\newline in terms of the basis functions
\cite{caboart}.

$|\varphi^{(r,\sigma_z)}_{\textbf{k}}(\vec{x},s)>=\sqrt{\frac{2}{N}}u^{\sigma_z}(s)\sum_{\vec{R}^{(r)}}\exp(i\textbf{k}\cdot\vec{R}^{(r)})\varphi_0(\vec{x}-\vec{R}^{(r)})$.\\

The formula (\ref{eqn_EnergAnis4}) can be wrote as follows:

\begin{equation}\label{eqn_EnergAnis3}
    E(\vec{n},\theta)=g\sum_{k,l}\sum_{r,\sigma_z}\sum_{r,\alpha_z}B_{r,\sigma_z}^{k,l}B_{t,\alpha_z}^{k,l}f_{\alpha_{z},\sigma_{z}}(\vec{n},\theta)g(r,t),
\end{equation}

\begin{equation*}
f_{\alpha_{z},\sigma_{z}}(\vec{n},\theta)=u^{\alpha_{z}}U^{+}\sigma_{z}Uu^{\sigma_z},
\end{equation*}

\begin{equation*}
g(r,t)=\frac{2}{N}\sum_{R^{r},R^{t}}e^{i\textbf{k}\cdot(\vec{R^{r}}-\vec{R^{t}})}\int d^{2}\vec{x}\varphi_{0}(\vec{x}-\vec{R^{t}})[E_{x}\hat{p}_{y}-E_{y}\hat{p}_{x}]\varphi_{0}(\vec{x}-\vec{R^{r}}).
\end{equation*}

In the nearest neighbor approximation, the term $g(r,t)$ take the following form:

\begin{equation}\label{eqn_orbrot3}
g(r,t)=\begin{cases}
          \int d^{2}\vec{x}\varphi_{0}(\vec{x})[E_{x}\hat{p}_{y}-E_{y}\hat{p}_{x}]\varphi_{0}(\vec{x}) & \text{if r=t} \\
          \sum^{1}_{n=-1}e^{ik_{x}np}\int_{V^{n}_{2}} d^{2}\vec{x}\varphi_{0}(\vec{x}+\vec{e_{x}}np)[E_{x}\hat{p}_{y}-E_{y}\hat{p}_{x}]\varphi_{0}(\vec{x})+ \\
          \exp^{ik_{y}np}\int_{V^{n}_{3}} d^{2}\vec{x}\varphi_{0}(\vec{x}+\vec{e_{y}}np)[E_{x}\hat{p}_{y}-E_{y}\hat{p}_{x}]\varphi_{0}(\vec{x}) & \text{if $r\neq t$}
              \end{cases}
\end{equation}

that can be expressed in more explicit form according to:

\begin{equation}\label{eqn_orbrot4}
g(r,t)=\begin{cases}
          A\int d^{2}\vec{x}e^{-\frac{\vec{x}^2}{2\ a^2}}[E_{y}x-E_{x}y]e^{-\frac{\vec{x}^2}{2\ a^2}} & \text{if r=t} \\
          \sum^{1}_{n=-1}e^{ik_{x}np}\int_{V^{n}_{2}} d^{2}\vec{x}e^{-\frac{|\vec{x}+\vec{e_{x}}np|^2}{2\ a^2}}[E_{y}x-E_{x}y]e^{-\frac{\vec{x}^2}{2\ a^2}}+ \\
          e^{ik_{y}np}\int_{V^{n}_{3}} d^{2}\vec{x}e^{-\frac{|\vec{x}+\vec{e_{y}}np|^2}{2\ a^2}}[E_{y}x-E_{x}y]e^{-\frac{\vec{x}^2}{2\ a^2}} & \text{if $r\neq t$}
              \end{cases}
\end{equation}

\noindent where $\varphi_{0}(\vec{x}) = \frac{1}{\sqrt{\pi a^2}}\exp(-\frac{\vec{x}^2}{2\ a^2})$ and $A=\frac{-\hbar i }{\pi a^4}$.

The integration regions are taken in what follows of fixed dimensions because
the Wannier orbitals are very much localized in the \ce{Cu} lattice centers.
$(a=0.25$ $p$, $p=3.8$ $A^{0})$:\\

$V_{1} = {(x,y)|-\frac{p}{2}<x,y<\frac{p}{2}}$,\\

$V^{+1}_{2} = {(x,y)|-\frac{p}{2}<x<\frac{p}{2},-\frac{3}{2}\frac{p}{2}<y<\frac{p}{2}}$,\\

$V^{-1}_{2} = {(x,y)|-\frac{p}{2}<x<\frac{p}{2},-\frac{p}{2}<y<\frac{3}{2}\frac{p}{2}}$,\\

$V^{+1}_{3} = {(x,y)|-\frac{3}{2}\frac{p}{2}<x<\frac{p}{2},-\frac{p}{2}<y<\frac{p}{2}}$,\\

$V^{-1}_{3} = {(x,y)|-\frac{p}{2}<x<\frac{3}{2}\frac{p}{2},-\frac{p}{2}<y<\frac{p}{2}}$.\\

\begin{figure}[ht]
\begin{center}
  \includegraphics[scale=.25]{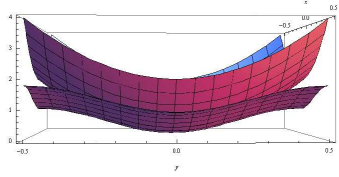}\\
\end{center}
  \caption{Parabolic potential (up) and effective potential (down) that coincide in the vicinity of the \ce{Cu} atoms. The unit of energy is $E=8.3$ $eV$ and the length unit $p=3.8$ $A^{0}$.}\label{fig_Potencialefectivo1}
\end{figure}

As the Wannier orbitals are two-dimensional Gaussian functions, they coincide
with the wavefunctions of the Harmonic two-dimensional oscillator. Therefore,
from the analysis already done, we can easily derive the effective potential
in the vicinity of the \ce{Cu} as a parabolic potential (figure
\ref{fig_Potencialefectivo1}). \vspace{1.5cm}

\begin{figure}[ht]
\begin{center}
  \includegraphics[scale=.25]{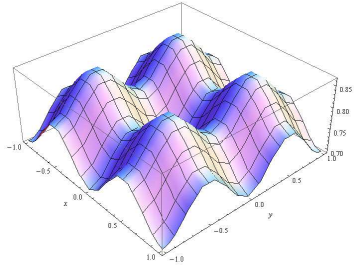}\\
\end{center}
  \caption{Effective potential with the order symmetry equal to 4 of the crystalline lattice, with parabolic form around the \ce{Cu} atoms, Gaussian Wannier orbitals. The unit of energy is $E=8.3$ $eV$ and the length unit $p=3.8$ $A^{0}$.}\label{fig_Potencialefectivo}
\end{figure}

The components $E_{x}$, $E_{y}$ of the electric field are derived form the
effective potential, see figure \ref{fig_Potencialefectivo}. They are odd
functions in the $x,y$ coordinates, a property that makes that the integral
factor $E_{y}x-E_{x}y$ is an odd function. As the integration regions are
symmetrical regarding, at least one of the coordinates, it follows that
$g(r,t)=0$.

This result evidences that the Gaussian Wannier orbitals in our small
overlapping approximation, predict a vanishing anisotropy energy. This means
that the first correction to the HF energy for the system, determined by the
spin-orbit interaction doesn't determine an easy magnetization axis. This
result is not coinciding with the observations, which indicates the existence an easy axis at zero doping and $\ T=0.$ \ \ However, this difference can be attributed to
various factors. One of them is the known absence of \ long range order in the
pure 2D systems. Another cause, can be the fact that the model slightly
simplifies the \ microscopic structure of the electron system by assuming it as
pure two dimensional and moreover, by considering Wannier orbitals not
resembling open the shell properties of the \ $Cu$ atoms.

The reduced form of $g(r,t)$ (\ref{eqn_orbrot3}) indicates that their value
only depends on the properties of symmetry of the electric field and the
orbitals in a vicinity of the \ce{Cu}, and not in the whole lattice. It points that
out is fundamental to more deeply formulate the starting basis of \ states for
the model to predict the magnetic properties.

\section{Anisotropy energy with Wannier orbitals being odd or even functions}

In this section, it is considered a modification of the orbitals of the model in
reference \cite{caboart,cabosimetria}. It is supposed that the Wannier
orbitals are three-dimensional. Their symmetry is assumed with the same
properties as the original problem. The Wannier obitals in general sense will
be denoted by $\psi(x_{1},x_{2},x_{3})$. It is supposed that they are even or
odd functions regarding the inversion of anyone of their coordinates, referred
to the \ce{Cu} atoms in that they are centered. In this case the spin-orbit
operator should have the usual form in three dimensions :

\begin{eqnarray}\label{eqn_SOexp1}
  \hat{U}_{SO}=-\frac{\hbar}{2m^{2}c^2}[\hat{\sigma_{x}}.(\partial_{y}V\hat{p}_{z}-\partial_{z}V\hat{p}_{y}) + \hat{\sigma_{y}}.(\partial_{z}V\hat{p}_{x}-\partial_{x}V\hat{p}_{z}) + \hat{\sigma_{z}}(\partial_{x}V\hat{p}_{y}-\partial_{y}\hat{p}_{x})].
\end{eqnarray}

From the definition of the anisotropy energy (\ref{eqn_EnergAnis}) and
(\ref{eqn_SOexp1}), it follows that it is a lineal combination of terms of the form:

\begin{eqnarray}\label{eqn_comblineal}
   E(\vec{n},\theta)=\sum^{3}_{i=1,j=1,i\neq j}(A_{ij}H_{ij0}+B_{ij+}H_{ij1+}+B_{ij-}H_{ij1-}+C_{ij+}H_{ij2+}+C_{ij-}H_{ij2-}),
\end{eqnarray}

\noindent where the $H_{ijk\pm}$ are of the following form:

$H_{ij0}=\int_{V_{0}}dx_{1}dx_{2}dx_{3}\psi(x_{1},x_{2},x_{3})E_{x_{i}}\hat{p}_{j}\psi(x_{1},x_{2},x_{3})$,\\

$H_{ij1\pm}=\int_{V_{1\mp}}dx_{1}dx_{2}dx_{3}\psi(x_{1}\pm p,x_{2},x_{3})E_{x_{i}}\hat{p}_{j}\psi(x_{1},x_{2}\pm p,x_{3})$,\\

$H_{ij2\pm}=\int_{V_{2\mp}}dx_{1}dx_{2}dx_{3}\psi(x_{1},x_{2}\pm p,x_{3})E_{x_{i}}\hat{p}_{j}\psi(x_{1}\pm p,x_{2},x_{3})$,\\

\noindent where the regions of integration $V_{1},V_{3x\mp},V_{2y\mp}$ can have in
general sense three possible spatial symmetries, when are defined by:\\

$V_{0} = {(x,y,z)|-\frac{p}{2}<x,y<\frac{p}{2}, -\frac{h}{2}<z<\frac{h}{2}}$,\\

$V_{1+} = {(x,y,z)|-\frac{p}{2}<x<\frac{3}{2}\frac{p}{2}}, -\frac{p}{2}<y<\frac{p}{2}, -\frac{h}{2}<z<\frac{h}{2}$,\\

$V_{1-} = {(x,y,z)|-\frac{3}{2}\frac{p}{2}<x<\frac{p}{2}}, -\frac{p}{2}<y<\frac{p}{2}, -\frac{h}{2}<z<\frac{h}{2}$,\\

$V_{2+} = {(x,y,z)|-\frac{p}{2}<x<\frac{p}{2}, -\frac{p}{2}<y<\frac{3}{2}\frac{p}{2}}, -\frac{h}{2}<z<\frac{h}{2}$\\,

$V_{2-} = {(x,y,z)|-\frac{p}{2}<x<\frac{p}{2}, -\frac{3}{2}\frac{p}{2}<y<\frac{p}{2}}, -\frac{h}{2}<z<\frac{h}{2}$.\\

Next it \ follows that:%

\begin{align}\label{eqn_demostrado}
  \int_{V_{0}}dx_{1}dx_{2}dx_{3}\psi(x_{1},x_{2},x_{3})E_{x_{i}}\hat{p}_{j}\psi(x_{1},x_{2},x_{3})=0,
\end{align}
\begin{align}\label{eqn_demostrado1}
  \int_{V_{1\mp}}dx_{1}dx_{2}dx_{3}\psi(x_{1}\pm p,x_{2},x_{3})E_{x_{i}}\hat{p}_{j}\psi(x_{1},x_{2}\pm p,x_{3})=0,
\end{align}
\begin{align}\label{eqn_demostrado2}
  \int_{V_{2\mp}}dx_{1}dx_{2}dx_{3}\psi(x_{1},x_{2}\pm p,x_{3})E_{x_{i}}\hat{p}_{j}\psi(x_{1}\pm p,x_{2},x_{3})=0.
\end{align}

It can be underlined that the two functions $p(x_{1},x_{2},x_{3})$ and
$t(x_{1},x_{2},x_{3})$ have the same parity regarding $x_{i}$ if both are even
or odd.

It is considered in our analysis that the effective potential near the \ce{Cu}
atoms of our model is even. Hence it follows that the fields $E_{x_{i}%
}=-\partial_{x_{i}}V$ are odd functions regarding $x_{i}$ and once fixed
$x_{i}$ , is even regarding the others coordinates $x_{j,k}$.

Let us show the (\ref{eqn_demostrado}-\ref{eqn_demostrado2}) properties now.

Proof of \ref{eqn_demostrado}.

\begin{enumerate}
\item The functions $\psi(x_{1},x_{2},x_{3})$ and $\partial_{x_{j}}\psi
(x_{1},x_{2},x_{3})$ has the same parity regarding $x_{i}$ and using the
initial premise about the fields, the integrand is an odd function concerning
the $x_{i}$ coordinate.

\item Let us use the result: If the integral of a function converges
absolutely, in an integration region in $R^{3}$ that has parallelepiped form,
then the iterate integrals can commute \ the integration order (Theorem of Fubini).

\item Using the previous result we concluded that $\int_{V_{0}}dx_{1}dx_{2}dx_{3}\psi(x_{1},x_{2},x_{3})E_{x_{i}}\hat{p}_{j}\psi(x_{1},x_{2},x_{3})=0$
\end{enumerate}

Proof of \ref{eqn_demostrado1}.

\begin{enumerate}
\item If $x_{i}=x_{1}$, then $\psi(x_{1},x_{2},x_{3})$ and $\partial_{x_{j}%
}\psi(x_{1}\pm p,x_{2},x_{3})$ have different parity regarding $x_{j}$ and
$E_{x_{i}}$ and continues being even regarding $x_{j}$, of the initial
premise, for what (\ref{eqn_demostrado1}) is kept.

\item If $x_{i}=x_{2,3}$, $\psi(x_{1},x_{2},x_{3})$ and $\partial_{x_{j}}%
\psi(x_{1}\pm p,x_{2},x_{3})$ have identical parity regarding $x_{i}$ and using
the initial premise of the fields, then the integrand is odd regard the
$x_{i}$ coordinate, for that (\ref{eqn_demostrado1}) is kept.

\item In both cases the before cited Fubini property was employed.
\end{enumerate}

Proof of \ref{eqn_demostrado2}

\begin{enumerate}
\item If $x_{i}=x_{2}$, it follows that $\psi(x_{1},x_{2},x_{3})$ and
$\partial_{x_{j}}\psi(x_{1},x_{2}\pm p,x_{3})$ have different parity regard
$x_{j},$ and $E_{x_{i}}$ continues been even regard $x_{j}$, of the initial
premise, for what it vanishes.

\item If $x_{i}=x_{1,3}$, $\psi(x_{1},x_{2},x_{3})$ and $\partial_{x_{j}}%
\psi(x_{1},x_{2}\pm p,x_{3})$ have identical parity regard $x_{i}$ and using
the initial premise of the fields, then has an odd integrand regard the
$x_{i}$ coordinate, and (\ref{eqn_demostrado2}) is kept.

\item In both cases the result of Fubini also was used.
\end{enumerate}

The results are summarized in the following form: If the Wannier orbitals that
we propose are even or odd functions regard the space coordinates and the
effective potential that it is assumed in the vecinity of the \ce{Cu} atoms,
is an even function regards the space coordinates, the anisotropy energy in
the model also vanishes and therefore an axis easy magnetization is not
predicted. Notice that the condition, that Wannier orbitals that we
propose are even or odd functions, is enough but not necessary condition for
the vanishing of the energy. This fits the possibility that using orbitals
that neither be even or odd, a non vanishing anisotropy energy could follows.

As it is known that the less bound electrons of the \ce{Cu2+} occupy the $d$
orbitals, and the electrons \ce {Cu2 +} considered into our model, are
representing the one partially filling the most energetic band in
$La_{2}CuO_{4},$ we can expect that the Wannier orbitals of our problem should
more closely resemble the $d_{x^{2}-y^{2}}$orbitals of $Cu$. These
$d_{x^{2}-y^{2}}$ orbitals are solutions of the hydrogen atom,. Then, we
could assume as the potential in the vicinities of the \ce{Cu} a Coulomb
potential alternatively as it was considered in the initial model. The Wannier
orbitals of the $d_{x^{2}-y^{2}}$ type are even functions regarding the space
coordinates. The components of the electric field $E_{x_{i}}$ are odd
functions because derived of the coulomb potential. Then we uses the previous
result to conclude that, if we suppose orbitals of the $d_{x^{2}-y^{2}}$ type then the model again will predicts anisotropy energy null and non easy axis of magnetization. \ However, the possibility of employing a general linear combination of the $d$ orbitals of $Cu$ in addition to a more realistic $3D$ character of the model by including multiple $CuO$ planes, opens the the opportunity to describe the observed antiferromagnetic anisotropy at low doping and temperatures.

\section{Spin-orbit operator in the Hartree-Fock approximation }

To study the influence of the spin-orbit operator in the Hartree-Fock approximation the operator is projected in the basis of the orbitals proposed in the model:

\begin{eqnarray}\label{EcuMatricial1}
[\textrm{H}_{\textbf{k}}+ \widetilde{\chi} \ (
\textrm{J}_{\textbf{k}}-\textrm{K}_{\textbf{k}}-\textrm{F}_{\textbf{k}})]\textbf{.}\textrm{B}^{\textbf{k},l}=
\widetilde{\varepsilon}_{l}(\textbf{k}) \
\textrm{S}_{\textbf{k}}\textbf{.}\textrm{B}^{\textbf{k},l}.
\end{eqnarray}

We can include the spin operator to obtain the next matrix equation

\begin{eqnarray}\label{EcuMatricial2}
[\textrm{H}_{\textbf{k}}+ \hat{\textrm{U}}^{SO}_{\textbf{k}}+ \widetilde{\chi} \ (
\textrm{J}_{\textbf{k}}-\textrm{K}_{\textbf{k}}-\textrm{F}_{\textbf{k}})]\textbf{.}\textrm{B}^{\textbf{k},l}=
\widetilde{\varepsilon}_{l}(\textbf{k}) \
\textrm{S}_{\textbf{k}}\textbf{.}\textrm{B}^{\textbf{k},l},
\end{eqnarray}

where the derivation of the term $\hat{\textrm{U}}^{SO}_{\textbf{k}}$ in the model is presented next:

\begin{eqnarray*}
  <\varphi^{(t,\alpha_{z})}_{\textbf{k}}(\vec{x},s)|\hat{U}_{SO}|\phi_{k,l}>&=&<\varphi^{(t,\alpha_{z})}_{\textbf{k}}(\vec{x},s)|A\hat{\sigma_{z}}(E_{x}\hat{p}_{y}-E_{y}\hat{p}_{x})\sum^{r,\sigma_z}B_{r,\sigma_z}^{k,l}|\varphi^{(r,\sigma_z)}_{\textbf{k}}>,  \\
  <\varphi^{(t,\alpha_{z})}_{\textbf{k}}(\vec{x},s)|\hat{U}_{SO}|\phi_{k,l}>&=&\sum^{r,\sigma_z}<\varphi^{(t,\alpha_{z})}_{\textbf{k}}(\vec{x},s)|A\hat{\sigma_{z}}(E_{x}\hat{p}_{y}-E_{y}\hat{p}_{x})|\varphi^{(r,\sigma_z)}_{\textbf{k}}>B_{r,\sigma_z}^{k,l},
\end{eqnarray*}

where $A=-\frac{\hbar}{2m^{2}c^2}$\\

\begin{equation*}
  \hat{U}^{SO}_{\textbf{k}} = A<\varphi^{(t,\alpha_{z})}_{\textbf{k}}(\vec{x},s)|\hat{\sigma_{z}}.(E_{x}\hat{p}_{y}-E_{y}\hat{p}_{x})|\varphi^{(r,\sigma_z)}_{\textbf{k}}>,
\end{equation*}

\begin{equation*}
  \hat{U}^{SO}_{\textbf{k}} = A\frac{2}{N}\sigma_{z}\delta_{\alpha_{z},\sigma_{z}}\sum^{\vec{R}^{(r)},\vec{R}^{(t)}}e^{i\vec{k}(\vec{R}^{(t)}-\vec{R}^{(r)})}\int\varphi_0(\vec{x}-\vec{R}^{(t)})(E_{x}\hat{p}_{y}-E_{y}\hat{p}_{x})\varphi_0(\vec{x}-\vec{R}^{(r)}).
\end{equation*}

In the near neighbor approximation the term $\hat{U}^{SO}_{\textbf{k}}$ take the following form:\\
\begin{equation*}
  \hat{U}^{SO}_{\textbf{k}} = A\sigma_{z}\delta_{\alpha_{z},\sigma_{z}}[\delta_{t,r}I_{1}+(\delta_{t-1,r}+\delta_{t,r-1})(e^{ik_{1}p}I_{-2}+e^{-ik_{1}p}I_{+2}+e^{ik_{2}p}I_{-3}+e^{-ik_{2}p}I_{+3})],
\end{equation*}\\

$I_{1}=\int_{V_{1}}d^{2}\vec{x}\varphi_0(\vec{x})(E_{x}\hat{p}_{y}-E_{y}\hat{p}_{x})\varphi_0(\vec{x})$,\\

$I_{\mp2}=\int_{V_{\mp3}}d^{2}\vec{x}\varphi_0(\vec{x}\mp p\vec{e}_{x})(E_{x}\hat{p}_{y}-E_{y}\hat{p}_{x})\varphi_0(\vec{x})$,\\

$I_{\mp3}=\int_{V_{\mp2}}d^{2}\vec{x}\varphi_0(\vec{x}\mp p\vec{e}_{y})(E_{x}\hat{p}_{y}-E_{y}\hat{p}_{x})\varphi_0(\vec{x})$.\\

For the same reasons that in section (anisotropy energy with orbitals of Wannier gaussian) the five integrals are null.\\

$I_{1}=I_{\mp2}=I_{\mp3}=0\Longrightarrow \hat{U}^{SO}_{\textbf{k}}=0$.\\

We can conclude that the spin-orbit operator is null in the matrix Hartree Fock representation, that implies no change in the Hartree Fock solution.

The same analysis could be done with the three dimensional Wannier orbitals and the result using the section(Anisotropy energy with orbitals of Wannier that are odd or even functions) would be the same as for bidimensional Gaussian Wannier orbitals.

\section{Summary}

\ \ We have studied the magnetic anisotropy properties of \ce{La2CuO4} at zero
temperature, in the context of a two-dimensional model of the layers \ce{CuO},
which Hartree Fock solutions were obtained in references\cite{caboart}. For
this purpose the spin-orbit interaction was considered as a perturbation to
the initial Hamiltonian of the system and the energy was evaluated as a function of the direction of the antiferromagnetic structure of the
material, when it is rotated. The basic conclusions of the work are:

1) The magnetic anisotropy energy strongly depends on the space dependence of
the orbitals of Wannier and the effective potential that the electrons feel in
the proximity of Cu.

2) For the Gaussian Wannier orbitals and their parabolic corresponding
effective potential employed in the model, it was obtained that the magnetic
anisotropy energy vanishes. Then, an easy axis of magnetization doesn't
appears and the antiferromagnetic structure doesn't align in any direction.

3) The use of more general Wannier orbitals, as can be Gaussian atomic
three-dimensional or orbital $d$ type, assuming that they are even or odd, all
produce vanishing vanishing anisotropy energy. This follows mainly because the
orbitals and the potentials are even or odd functions with regard the coordinates.

4) The approach employed, that only considers non null overlapping among
orbitals that are centered in near neighbors, is central in the previous conclusions.

5) The fact that the lattice of the \ce{Cu} atoms is squared, strongly
influences these results of \ zero magnetic anisotropy energy.

6) Wannier orbitals that are not spatially even neither odd, can predict non
zero values of the magnetic anisotropy energy.

\end{document}